\documentclass[twocolumn,amsmath,amssymb,nofootinbib,prd,aps,10pt,floatfix]{revtex4-2}
\usepackage[utf8]{inputenc}
\usepackage{graphicx, float,amsmath, amsmath, amssymb}

\usepackage{amsfonts}
\usepackage{bm}
\usepackage{dcolumn}
\usepackage[braket]{qcircuit}
\usepackage[export]{adjustbox}
\usepackage{multirow}
\usepackage{color}
\usepackage{braket}
\usepackage{caption}
\usepackage{subcaption}
\usepackage{tensor}
\usepackage{listings}
\usepackage{xcolor}
\usepackage{amsmath}
\usepackage{amsfonts}
\usepackage{float}
\usepackage{bbold}
\usepackage{mathtools}
\usepackage{tensor}
\usepackage{tikz}
\usetikzlibrary{positioning}

\definecolor{codegreen}{rgb}{0,0.6,0}
\definecolor{codegray}{rgb}{0.5,0.5,0.5}
\definecolor{codepurple}{rgb}{0.58,0,0.82}
\definecolor{backcolour}{rgb}{0.95,0.95,0.92}

\lstdefinestyle{mystyle}{
    backgroundcolor=\color{backcolour},   
    commentstyle=\color{codegreen},
    keywordstyle=\color{magenta},
    numberstyle=\tiny\color{codegray},
    stringstyle=\color{codepurple},
    basicstyle=\ttfamily\footnotesize,
    breakatwhitespace=false,         
    breaklines=true,                 
    captionpos=b,                    
    keepspaces=true,                 
    numbers=left,                    
    numbersep=5pt,                  
    showspaces=false,                
    showstringspaces=false,
    showtabs=false,                  
    tabsize=2
}

\usepackage{hyperref}

\usepackage{tikz}
\usetikzlibrary{decorations.markings}
\usetikzlibrary{calc}

\tikzset{
    ingoing/.style={decoration={markings, mark=at position 0.5 with {\arrow{<}}}, postaction=decorate},
    outgoing/.style={decoration={markings, mark=at position 0.5 with {\arrow{>}}}, postaction=decorate}
}

\newcommand{\dket}[1]{| #1 \rangle\rangle}

\newcommand{\mybra}[2]{\prescript{}{#2}{\bra{#1}}}

\begin{document}

\title{Spin networks of quantum channels}
\author{Bartosz Grygielski$^{1,2}$}
\email{bartosz.grygielski@doctoral.uj.edu.pl}
\author{Jakub Mielczarek$^{1}$}
\email{jakub.mielczarek@uj.edu.pl}
\affiliation{
$^{1}$Institute of Theoretical Physics, Jagiellonian University, 
{\L}ojasiewicza 11, 30-348 Cracow, Poland\\
$^{2}$Doctoral School of Exact and Natural Sciences, Jagiellonian University, 
{\L}ojasiewicza 11, 30-348 Cracow, Poland}
\date{\today}

\begin{abstract}
Spin networks in Loop Quantum Gravity are traditionally described by unitary 
holonomies corresponding to noiseless transformations. In this work, we extend 
this framework to incorporate general quantum channels that model effects 
of environment, which can become significant at the Planck scale. Specifically, 
we demonstrate that the transformation properties of Kraus operators, which define 
completely positive trace-preserving (CPTP) maps, are consistent with the gauge 
invariance of spin networks. This enables the introduction of generalized spin 
network states that can be expressed in terms of the Kraus operators. Furthermore,
the associated notion of an inner product is proposed, allowing for introduction  
of the Hilbert space. We illustrate these constructions with examples involving a 
Wilson loop and a dipole network. 
\end{abstract}

\maketitle

\section{Introduction}
\label{Sec.Introduction}

The concept of spin networks goes back to the seminal 
work of Penrose in the 1970s, where they were introduced 
as a combinatorial description of space \cite{Penrose1971NegativeDimensionalTensors}. These ideas were 
later revived in the late 80s, leading to the framework of 
loop quantum gravity \cite{Rovelli:1987df,Rovelli:1997yv,Ashtekar:2004eh}, 
which was being formulated at that time.

This approach is based on the Ashtekar-Sen variables \cite{Sen:1982qb,PhysRevLett.57.2244}, which 
reveal that General Relativity can be reformulated as an
$SU(2)$ non-Abelian gauge field theory \cite{Ashtekar:1987gu}. 
The key difference from ordinary Yang-Mills theories lies 
in the highly-nonlinear structure of the gravitational Hamiltonian, 
and in the fact that gravity is a purely constrained system.

The $SU(2)$ gauge symmetry naturally gives rise to spin 
networks, which provide gauge-invariant states of quantum
geometry. Moreover, the action of geometric operators becomes
particularly transparent when expressed in the spin-network
basis.

More recently, increasing attention has been devoted 
to the quantum-information-theoretic properties of spin 
networks \cite{bianchi2023loop,Oriti:2025uad}. In particular, their 
entanglement structure has been studied extensively \cite{Mielczarek:2019srn,livine2018intertwiner}. 
The studies also lead to the first quantum simulations 
of spin networks \cite{Czelusta:2020ryq,Li:2017gvt,Czelusta:2023dij,Mielczarek:2018jsh} 
and unveiled relations between spin networks 
and tensor networks. This, in particular, turned out to 
be relevant in the context of understanding holographic properties
in LQG \cite{Han:2016xmb,Czelusta:2024tvz}

Furthermore, it has been shown that the links of spin 
networks are naturally associated with maximally entangled 
states, which act as a form of ``glue'' joining the ``atoms 
of space'' \cite{Baytas:2018wjd, Bianchi:2018fmq, Czelusta:2024tvz}.
This connection arises because the links of spin networks,
which are associated with $SU(2)$ holonomies, can be
represented as maximally entangled states through the 
Choi-Jamio{\l}kowski isomorphism \cite{Choi:1975nug,JAMIOLKOWSKI1972275}.
Importantly, this construction can, in principle, be extended 
to non-unitary quantum channels \cite{kraus1971general,holevo1998quantum,rivas2012open}, 
thereby generalizing the notion of unitary holonomies.

The goal of this article is to explore this direction and to 
study possible generalizations of spin networks beyond unitary 
holonomies. Such a framework would allow one to incorporate 
environmental effects to which quantum gravitational degrees 
of freedom may be sensitive, and it opens a pathway toward 
analyzing decoherence processes in quantum gravity.

The organization of the article is as follows. In Sec. \ref{Sec:SpinNetworks} 
we review the basic properties of spin networks. In Sec. \ref{Sec:QuantumChannels} we introduce 
the notion of quantum channels and discuss their properties, 
in particular in the context of incorporating environmental 
degrees of freedom. In Sec. \ref{Sec:Holonomies} we present a basic 
construction that extends holonomies to quantum channels and verify the transformation 
properties of the resulting objects. In Sec. \ref{Sec:WilsonLoop} we show that this 
construction allows for a generalization of Wilson loops while 
preserving gauge invariance. Building on this result, in Sec. \ref{Sec:GeneralizedNetworks}
we construct a generalized notion of spin-network states and the 
associated Hilbert space. The results are illustrated by the 
discussion of the normalization of the quantum-channel-based 
Wilson loop state and the dipole spin network. The framework is further 
explored in Sec. \ref{Sec:GeneralConstruction}, where possible generalizations at the level 
of superpositions of spin-network states are discussed. Finally, 
in Sec. \ref{Sec.Summary} we summarize our results and outline directions for 
future research. Additional properties of the Kraus operators are 
summarized in the Appendix.

\section{Spin networks}
\label{Sec:SpinNetworks}

In gauge theories, translations in space influence the 
internal degrees of freedom of a particle. This can be 
described by holonomies (Wilson lines), defined as:
\begin{equation}
    h_\gamma[A]\coloneqq\mathcal{P}\exp-\int_\gamma A,
\end{equation}
where \(A=A_adx^a = A^i_a\tau_i dx^a \) is the gauge field 
(connection 1-form) of corresponding theory, and \(\mathcal{P}\) 
is the path-ordering operator, and \(\gamma\) is the path traversed 
by the particle, \(\gamma\: :\:[0,1]\rightarrow\Sigma\), with \(\Sigma\) being 
the hypersurface of constant time, 
and we denote \(\gamma(0)=s\) and \(\gamma(1)=t\), referring 
to the \textit{source} and the \textit{target} spaces, \(\mathcal{H}_s\) 
and \(\mathcal{H}_t\), respectively. The $\tau_i$, with $i=1,2,3$,
are $SU(2)$ generators satisfying the algebra 
$[\tau_i,\tau_j]=\epsilon_{ijk}\tau_k$.

Under the gauge transformation: 
\begin{equation}
A_a=A^{i}_a \tau_i  \rightarrow A'_a= UA_a U^{\dagger}+U^{\dagger}\partial_a U, 
\end{equation}
where $U\in SU(2)$, the holonomy transforms as:
\begin{equation}
h_{\gamma}[A] \rightarrow h_{\gamma}[A'] = U_t  h_{\gamma}[A]  U_s^{\dagger},
\label{eq_holonomy_transformation}
\end{equation}
where $U_s:=U(s)$ and $U_t:=U(t)$.

Following the transformation property  (\ref{eq_holonomy_transformation}) 
one can notice that the holonomy is a unitary map $\hat{h}\in \mathcal{H}_t
\otimes  \mathcal{H}_s^*$, so its right-handed action, on a ket state, 
is: 
\begin{equation}
\hat{h}_R : \mathcal{H}_s \rightarrow \mathcal{H}_t.
\end{equation}

The transformation rule (\ref{eq_holonomy_transformation})
makes the holonomy a handy starting point for constructing 
gauge-invariant objects. The basic construction is the Wilson loop:
\begin{equation}
W[h_{\gamma}[A]] := {\rm tr}(h_{\gamma}[A]),
\end{equation}
which is explicitly a gauge-invariant object. 

The construction can be generalized to 
the spin-network states, which are 
square integrable functions of the form:
\begin{equation}
    \Psi_{\Gamma}[A] :=\psi(h_{e_1}[A],\dots,h_{e_E}[A]).
\end{equation}
Introducing the Ashtekar-Lewandowski measure $d\mu_{AL}$ 
\cite{Ashtekar:1994wa,Ashtekar:1994mh}, and the 
scalar product:
\begin{equation}
\braket{\Psi_{\Gamma}|\Phi_{\Gamma'}} :=  \int d\mu_{AL} \overline{\Psi_{\Gamma}[A]} \Phi_{\Gamma'}[A],  
\end{equation}
the functions form a Hilbert space. Here, $d\mu_{AL}= 
\Pi_{e \in \Gamma \cap \Gamma'}dg_e$, where $dg$ is the normalized 
Haar measure on $SU(2)$, such that $\int_{SU(2)}dg=1$.

The gauge-invariance is then imposed by virtue 
of the Gauss constraint:
\begin{equation}
 \sum_{i } \hat{F}_{S_i} \ket{\Psi} = 0, 
\end{equation}
where $\hat{F}_{S_i}$ are the flux operators corresponding
to the surfaces $S_i$ being adjacent to the links for a given
node. As a consequence, gauge-invariant states are obtained
by contracting the representation spaces carried by the incident 
links with an invariant tensor (an intertwiner), which ensures 
that the node is gauge-invariant.

The fluxes are classically introduced as:
\begin{equation}
F^j_S[E] := \int_S \epsilon_{abc} E^a_j dx^b \wedge dx^c,
\label{DefFlux}
\end{equation}
where $E^a_i$ is the $SU(2)$ ``electric field'', canonically
conjugate to the gauge field $A^i_a$, such that the holonomy-flux
algebra is satisfied:
\begin{equation} \label{Eq: H-F alg}
\{h_{\gamma}[A], F^j_S [E] \} = -\iota(\gamma,S)
h_{\gamma_1}[A]\tau^j h_{\gamma_2}[A].
\end{equation}
Here, the path $\gamma \subset \Sigma$ is decomposed as
$\gamma=\gamma_1\cup\gamma_2$ at the intersection point 
with the surface $S$, and the value of $\iota(\gamma,S)=0,\pm 1$ 
depends on the relative orientation and intersection properties 
of $\gamma$ and $S$: it is zero if $\gamma$ does not intersect $S$, 
and $\pm 1$ if $\gamma$ intersects $S$, with the sign determined 
by whether the oriented tangent vector of $\gamma$ at the intersection 
point is aligned or anti-aligned with the chosen orientation of $S$.

\section{Quantum channels}
\label{Sec:QuantumChannels}

Considering open quantum systems requires describing 
quantum systems with density matrices \(\hat{\rho}\). 
Any density matrix can be viewed as a positive-semidefinite linear operator from 
a Hilbert space to itself (an automorphism):
\begin{equation}
    \hat{\rho}\in\mathcal{B}(\mathcal{H})_{\geq0},
\end{equation}
\begin{equation}
    \hat{\rho}:\:\mathcal{H}\rightarrow\mathcal{H},
\end{equation}
where \(\mathcal{B}(\mathcal{H})_{\geq0}\) denotes positive-semidefinite 
cone of the space of bounded linear operators over the space \(\mathcal{H}\). 
Positive semidefiniteness of, e.g., operator \(\hat{\rho}\) over space 
\(\mathcal{H}\) will be denoted as \(\hat{\rho}\geq0\), and by definition means:
\begin{equation}
    \hat{\rho}\geq0\iff\bra{\psi}\hat{\rho}\ket{\psi}\geq{0}\;\;\forall \ket{\psi}\in\mathcal{H}.
\end{equation}
Evolution (not necessarily temporal) of density matrices maps:
\begin{equation}
    \hat{\Lambda}:\:\mathcal{B}(\mathcal{H}_s)_{\geq0}\rightarrow\mathcal{B}(\mathcal{H}_t)_{\geq0},
\end{equation}
where subscripts \(s\) and \(t\) denote source and target spaces, 
respectively. At the level of density matrices, this is denoted:
\begin{equation}
    \hat{\rho}_t=\hat{\Lambda}(\hat{\rho}_s).
\end{equation}

\par Quantum channels are generalizations of unitary maps, 
that encompass noise from the environment of the considered system. 
By definition, they are given by linear, completely positive (CP) 
and trace preserving (TP) maps - together - CPTP maps. Positivity 
of a map ensures that a density matrix that is \(\hat{\rho}\geq0\) gets 
mapped to another density matrix that is positive semidefinite, 
thus retaining a probabilistic interpretation of quantum mechanics. 
Complete positivity, however, lets us apply this condition with 
the presence of an environment, which remains unchanged, and by 
definition means:
\begin{equation}
    (\hat{\Lambda}\otimes\hat{\mathbb{I}}_n)\geq0,
\end{equation}
for any \(n\in\mathbb{N}\) - dimension of environment. 
Trace preservation guarantees that no probability is lost.
\par Due to Choi's theorem on completely positive maps 
\cite{Choi_1975}, any such map can be written via Kraus operators:
\begin{equation}
    \hat{\Lambda}(\hat{\rho}):=\sum_i\hat{K}_i\hat{\rho} \hat{K}_i^\dagger,
\end{equation}
where \(\hat{K}_i\) are the Kraus operators describing the effective evolution of the system. For their derivation from microscopic dynamics and interpretation, see Appendix. It is easy to show that they fulfill the CP condition. Consider a joint density matrix of some system \(S\) and environment \(E\), \(\hat{\rho}_{SE}\geq0\). Due to positive-semidefiniteness, this means that:
\begin{equation}
    \bra{\psi}\hat{\rho}_{SE}\ket{\psi}\geq0,
\end{equation}
where \(\ket{\psi}\) is any state in the joint space. An evolved state is going to be: 
\begin{equation}
    \hat{\rho}_{SE}'=\left(\hat{\Lambda}\otimes\hat{\mathbb{I}}\right)(\hat{\rho}_{SE})=\sum_i(\hat{K}_i\otimes \hat{\mathbb{I}})\hat{\rho}_{SE}( \hat{K}_i^\dagger\otimes\hat{\mathbb{I}}).
\end{equation}
Let us then consider the expected value of the evolved state:
\begin{align}
    \begin{split}
        \bra{\psi}\hat{\rho}_{SE}'\ket{\psi}&=\sum_i\underbrace{\bra{\psi}(\hat{K}_i\otimes \hat{\mathbb{I}})}_{\bra{\psi_i}}\hat{\rho}_{SE}\underbrace{(\hat{K}_i^\dagger\otimes \hat{\mathbb{I}})\ket{\psi}}_{\ket{\psi_i}}
       \\ &=\sum_i\underbrace{\bra{\psi_i}\hat{\rho}_{SE}\ket{\psi_i}}_{\geq0}\geq0,
    \end{split}
\end{align}
where we have used the fact that \((\hat{K}_i\otimes\hat{\mathbb{I}})\ket{\psi}\) yields another vector \(\ket{\psi_i}\) in the same Hilbert space. Trace preservation is enforced by the condition:
\begin{equation}
    \sum_i \hat{K}_i^\dagger \hat{K}_i=\hat{\mathbb{I}}.
    \label{eq_kraus_trace_preservation}
\end{equation}
This is again easy to see by explicitly tracing an evolved density matrix:
\begin{align}
    \text{tr}(\hat{\Lambda}(\hat{\rho})) & = \text{tr}\left(\sum_i\hat{K}_i\hat{\rho} \hat{K}_i^\dagger\right) \nonumber  \\
     &=\text{tr}\left(\sum_i\hat{K}_i^\dagger \hat{K}_i\hat{\rho}\right)=\text{tr}(\hat{\rho}).
\end{align}
Let us stress again that an evolution operator is a quantum channel if, and only if, it can be written in a Kraus form. The most important features of this description are listed below:
\begin{itemize}
    \item \(\hat{\Lambda}\otimes\hat{\mathbb{I}}_n\geq0\quad\forall n\in\mathbb{N}\),
    \item \(\hat{\Lambda}\) - CP \(\iff\) it can be written in a Kraus form \(\hat{\Lambda}(\hat{\rho})=\sum_i\hat{K}_i\hat{\rho} \hat{K}_i^\dagger\),
    \item If \(\hat{\Lambda}\) - Trace preserving: \( {\rm tr}(\hat{\Lambda}(\hat{\rho}))={\rm tr}(\hat{\rho})\), then it is necessary and sufficient that \(\sum_i\hat{K}^\dagger_i\hat{K}_i=\hat{\mathbb{I}}\).
\end{itemize}

\subsection{Purity of quantum channel-evolved states}

Evolution via a quantum channel changes the purity of a state. 
Specifically, we can consider the action of a quantum channel 
on a pure state \(\hat{\rho}=\ket{\psi}\bra{\psi}\) and show 
that it results in a mixed state. This is best done by explicitly 
acting on \(\hat{\rho}\) with a general form of Kraus evolution:
\begin{equation}
\hat{\rho}'=\sum_i\underbrace{(\hat{K}_i\otimes\hat{\mathbb{I}})\ket{\psi}}_{\coloneq\ket{\psi_i}}\underbrace{\bra{\psi}(\hat{K}_i^\dagger\otimes\hat{\mathbb{I}})}_{\bra{\psi_i}}=\sum_i\ket{\psi_i}\bra{\psi_i},
\end{equation}
and observing that this is a sum of pure states, which in general will be mixed. It would be pure only in the case where for all \(\hat{K}_i\), \(\hat{K}_i\otimes\hat{\mathbb{I}}\) map \(\ket{\psi}\) to some \(\hat{\Lambda}_i\ket{\phi}\), and \(\sum_i\hat{\Lambda}_i=1\). That, however, would mean that the rank of the Choi matrix is 1, which implies unitary evolution, making the Kraus evolution description redundant for that case. This fact is important, e.g., in adapting the Choi-Jamiołkowski isomorphism to quantum channels.

\subsection{Bit flip model of the environment}
A simple example of evolution through a quantum channel is the bit flip model. It corresponds to identity evolution \eqref{eq_bit_flip_kraus_0} with probability \(p\) of interaction with the environment in such a way that the considered bit is flipped through \eqref{eq_bit_flip_kraus_1}. In computational basis, the Kraus operators corresponding to this model are:
\begin{equation}
    \hat{K}_0 =\hat{\mathbb{I}} \sqrt{1-p} = \sqrt{1-p}\begin{pmatrix}
    1 & 0\\
    0 & 1 \\
    \end{pmatrix},
    \label{eq_bit_flip_kraus_0}
\end{equation}
\begin{equation}
    \hat{K}_1 =\hat{\sigma}_x\sqrt{p} =  \sqrt{p} \begin{pmatrix}
    0 & 1\\
    1 & 0 \\
    \end{pmatrix}. 
    \label{eq_bit_flip_kraus_1}
\end{equation}
These will be used throughout this article to demonstrate some of our findings.

\section{Holonomies as quantum channels}
\label{Sec:Holonomies}

\subsection{Generalizing the theory}
When constructing the generalized theory, three conditions are demanded to be fulfilled:
\begin{enumerate}
    \item The unitary case can easily be described by the more general theory.
    \item Gauge transformation property of the holonomy \\\( \hat{h}'=\hat{U}_t \hat{h}\hat{U}^\dagger_s\) is preserved.
    \item The description in invariant under Kraus operator mixing \(\hat{K}'_i=\sum_lU_{i}^{l}\hat{K}_l\).
\end{enumerate}

The first condition makes the new theory easily comparable with already 
established results. The second is a necessity for describing gauge theory 
on a spin network. Finally, the third condition ensures that the description 
is unique and does not depend on the choice of basis in the Kraus operator space. 

\par Following the usual treatment of quantum open systems, we assume that at the kinematical level, which will be a ground for initial conditions for a
dynamical case, our system is not correlated with its environment, i.e., any 
density matrix \(\hat{\rho}^{SE}\) on the joint space \(\mathcal{H}_{\rm kin}=\mathcal{H}_S\otimes\mathcal{H}_E\) can be written as \(\hat{\rho}^{SE}=\hat{\rho}\otimes\ket{0}_{EE}\bra{0}\), which lets us define 
Kraus operators between the nodes. System's space \(\mathcal{H}_S\) remains 
covariant under gauge transformations. We assume that if we had full 
information about the environment could be described by unitary holonomies:
\begin{equation}
    \hat{\rho}^{SE}_t=\hat{h}\hat{\rho}^{SE}_s\hat{h}^\dagger,
\end{equation}
but when considering a case where we only have access to the system's degrees 
of freedom and observables, we want to average over the environmental degrees 
of freedom, which is done by taking a partial trace over this part of join Hilbert space. Then we get:
\begin{align}
    \hat{\rho}_t&={\rm tr}_E(\hat{h}\hat{\rho}^{SE}_s\hat{h}^\dagger) \nonumber \\
    &=\sum_i\mybra{i}{E}\hat{h}\Big(\hat{\rho}_s\otimes \ket{0}_{EE}\bra{0}\Big)\hat{h}^\dagger\ket{i}_E,
\end{align}
and acting with holonomy on state \(\ket{0}_E\), 
we can define the Kraus operators:
\begin{equation}
    \sum_i\mybra{i}{E}\hat{h}\ket{0}_E\hat{\rho}_s\mybra{0}{E}\hat{h}^\dagger\ket{i}_E\coloneq\sum_i \hat{K}_i\hat{\rho}_s \hat{K}_i^\dagger,
    \label{eq_Kraus_definition}
\end{equation}
thus converting to an effective description of an open system.

\subsection{Vectorization}

For practical purposes, it will be convenient to work in a vectorized picture for this approach. It explicitly preserves invariance under gauge transformations and under unitary Kraus operator mixing, and furthermore resembles the unitary picture of spin networks. Formally, vectorization maps an operator \(\hat{A}\in\mathcal{H}\otimes\mathcal{H}^*\rightarrow\mathrm{vec}(\hat{A})\in\mathcal{H}\otimes\mathcal{H}\). Vectorized operators will be denoted by a double ket: \(\mathrm{vec}(\hat{A})\coloneq\dket{\hat{A}}\). Vectorized operators belong to the \emph{Liouville space} \(\mathcal{L}\), which, under admitting a Hilbert-Schmidt product, is a Hilbert space.
\par Explicit vectorization on matrices can be realized in various ways, but here we will utilize column stacking, such that for a matrix \(\hat{A}=\begin{pmatrix}a & b \\ c & d\end{pmatrix}\), its vectorization will be given by:
\begin{equation}
    \dket{\hat{A}}=\begin{pmatrix} a \\ c \\ b \\ d\end{pmatrix}.
\end{equation}
It is important to note that this procedure is linear, which follows from the fact that the addition of matrices is performed element-wise. Another property arises when vectorizing a product of three operators, then one can represent the resulting vector as a linear operator constructed from two of the operators, acting on a third, vectorized, operator:
\begin{equation}
    \mathrm{vec}(\hat{A}\hat{B}\hat{C})=(\hat{C}^\mathrm{T}\otimes \hat{A})\mathrm{vec}(\hat{B}).
\end{equation}
These two properties let us write any quantum channel in the vectorized form as: 
\begin{align}
    \begin{split}
        \hat{\rho}_t&=\sum_i\hat{K}_i\hat{\rho}_s \hat{K}_i^\dagger,\\
        &\downarrow
        \\
        \dket{\hat{\rho}_t}&=\left(\sum_i\hat{K}_i^*\otimes \hat{K}_i\right)\dket{\hat{\rho}_s}\coloneq\hat{\Lambda}\dket{\hat{\rho}_s},
    \end{split}
\end{align}
where we define a linear operator \(\hat{\Lambda}=\sum_i \hat{K}_i^*\otimes \hat{K}_i\) 
that fully describes the action of the quantum channel on the source state, 
and does not refer to the state it acts on in the description of evolution. 
This also guarantees that there is no redundancy in the description, which 
is a possibility in formulating the evolution by Kraus operators, as described 
in the Appendix.

For the bit flip model \eqref{eq_bit_flip_kraus_0}, \eqref{eq_bit_flip_kraus_1} described in an earlier section, the operator becomes:
\begin{equation}
    \hat{\Lambda} =
    \begin{pmatrix}
        1-p & 0 & 0 & p \\
        0 & 1-p & p & 0 \\
        0 & p & 1-p & 0 \\
        p & 0 & 0 & 1-p
    \end{pmatrix},
\end{equation}
which explicitly describes the action of this channel on a generally mixed qubit state.

\subsection{Spin networks of quantum channels}

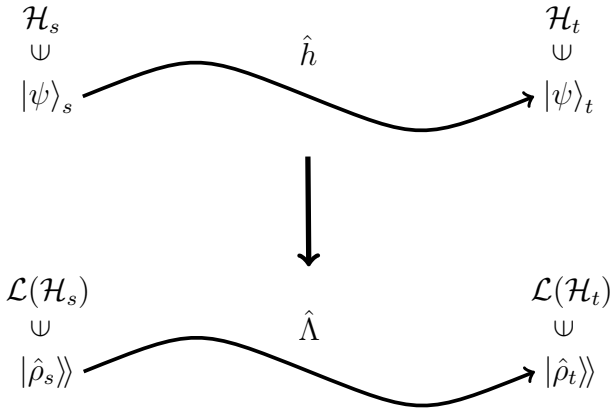
\begin{figure}[ht!]
    \centering
    \usetikzlibrary{positioning,decorations.pathmorphing}
    \begin{tikzpicture}[
      every node/.style={font=\large},
      map/.style={->, line width=1.4pt},
      lift/.style={->, line width=2pt}
    ]
    \node (Hs) {$\ket{\psi}_{s}$};
    \node (Ht) [right=6cm of Hs] {$\ket{\psi}_{t}$};
    \node (HsIn) [above=0.25cm of Hs, xshift=0.15cm, rotate=90] {$\in$};
    \node (HtIn) [above=0.25cm of Ht, xshift=0.15cm, rotate=90] {$\in$};
    \node (HsSpace) [above=0.4cm of Hs, xshift=-0.07cm] {$\mathcal{H}_{s}$};
    \node (HtSpace) [above=0.4cm of Ht, xshift=-0.07cm] {$\mathcal{H}_{t}$};
    \draw[map]
      (Hs.east) 
      .. controls ($(Hs.east) + (1.5cm,0.6cm)$) and ($(Hs.east)!0.5!(Ht.west) + (-1.5cm,0.6cm)$) .. ($(Hs.east)!0.5!(Ht.west)$)
      .. controls ($(Hs.east)!0.5!(Ht.west) + (1.5cm,-0.6cm)$) and ($(Ht.west) + (-1.5cm,-0.6cm)$) .. (Ht.west);
    \node[above=0.3cm] at ($(Hs.east)!0.5!(Ht.west)$) {$\hat{h}$};
    \node (Ls) [below=3cm of Hs] {$\ket{\hat{\rho}_{s}}\!\rangle$};
    \node (Lt) [right=6cm of Ls] {$\ket{\hat{\rho}_{t}}\!\rangle$};
    \node (LsIn) [above=0.25cm of Ls, xshift=0.15cm, rotate=90] {$\in$};
    \node (LtIn) [above=0.25cm of Lt, xshift=0.15cm, rotate=90] {$\in$};
    \node (LsSpace) [above=0.4cm of Ls, xshift=0.015cm] {$\mathcal{L}(\mathcal{H}_{s})$};
    \node (LtSpace) [above=0.4cm of Lt, xshift=0.02cm] {$\mathcal{L}(\mathcal{H}_{t})$};
    \draw[map]
      (Ls.east)
      .. controls ($(Ls.east) + (1.5cm,0.6cm)$) and ($(Ls.east)!0.5!(Lt.west) + (-1.5cm,0.6cm)$) .. ($(Ls.east)!0.5!(Lt.west)$)
      .. controls ($(Ls.east)!0.5!(Lt.west) + (1.5cm,-0.6cm)$) and ($(Lt.west) + (-1.5cm,-0.6cm)$) .. (Lt.west);
    \node[above=0.3cm] at ($(Ls.east)!0.5!(Lt.west)$) {$\hat{\Lambda}$};
    \draw[lift]
      ($(Hs)!0.5!(Ht)+(0,-0.8)$) -- ($(Ls)!0.5!(Lt)+(0,1.4)$);
    \end{tikzpicture}
    \caption{Pictorial representation of the extention introduced in this work.}
    \label{Fig:unitarychannelcomparison}
\end{figure}

Concepts described in the previous sections can be used to build 
a spin network consisting of density matrices and quantum channels 
instead of state vectors and unitary holonomies, as shown in Fig. 
\ref{Fig:unitarychannelcomparison}.

Here, we will discuss that such reformulation fully resembles the 
original theory of unitary spin networks and that it fulfills our requirements. 
The usual spin networks consist of the vertices of that are elements 
of a Hilbert space, and edges carrying unitary holonomies. 
Here, on the vertices we have maps over said Hilbert space that 
are connected by non-unitary, but still linear, quantum channels. Thus, we are 
solely loosening one of the conditions for the  construction of spin networks, while
maintaining conceptual simplicity. This description does not require a shift in
perspective while constructing physical setups or modelling noise. A unitary 
holonomy can be reconstructed from admitting only one Kraus operator, in which case 
\(\hat{\Lambda}=U^*\otimes U\), and the property \(\hat{\Lambda}\hat{\Lambda}^\dagger=\hat{\Lambda}^\dagger\hat{\Lambda}=\hat{\mathbb{I}}\) is maintained. 

Under gauge transformations, a general \(\hat{\Lambda}\) transforms as:
\begin{align}
    \begin{split}
        \hat{\Lambda}'&=\sum_i\hat{K}_i'^*\otimes \hat{K}_i'=\sum_i\hat{U}_t^*\hat{K}_i^*\hat{U}_s^\mathrm{T}\otimes \hat{U}_t\hat{K}_i\hat{U}_s^\dagger\\
        &=(\hat{U}_t^*\otimes \hat{U}_t)\left(\sum_i\hat{K}_i^*\otimes \hat{K}_i\right)(\hat{U}_s^\mathrm{T}\otimes \hat{U}_s^\dagger)\\
        &\coloneq\hat{\hat{\mathcal{U}}}_t\left(\sum_i\hat{K}_i^*\otimes \hat{K}_i\right)\hat{\mathcal{U}}_s^\dagger=\hat{\mathcal{U}}_t\hat{\Lambda}\hat{\mathcal{U}}_s^\dagger,
    \end{split}
    \label{eq_lambda_gauge_invariance}
\end{align}
where we have defined unitary operators \(\hat{\mathcal{U}}_{s/t}\). Derivation of gauge transformation rules for individual Kraus operators can be found in the Appendix. This is again fully analogous to the unitary case and preserves the required holonomy transformation rule under gauge transformations, albeit in the extended space. The last requirement we proposed was invariance under change of basis in the Kraus operator space \(\hat{K}_i'=\sum_l U_{il}\hat{K_l}\), which is easily shown to be fulfilled:
\begin{align}
    \begin{split}
        \hat{\Lambda}'&=\sum_i\hat{K}_i'^*\otimes \hat{K}_i'\\
        &=\sum_{ilk}\tensor{U}{^*_i^l}\hat{K}^*_l\otimes \tensor{U}{_i^k}\hat{K}_k=\sum_{ilk}\tensor{U}{^*_i^l}\tensor{U}{_i^k}\hat{K_l}^*\otimes \hat{K}_k\\
        &=\sum_{kl}\delta^{lk}\hat{K_l}^*\otimes \hat{K}_k=\sum_l\hat{K_l}^*\otimes \hat{K_l}=\hat{\Lambda}.
    \end{split}
\end{align}
\par Thus, we have constructed a theory that uses identical formalism as spin networks with unitary holonomies, but extends utility to quantum channels. This is at an unavoidable price of squaring the dimension of Hilbert spaces involved due to the inclusion of the environment and the tensor product construction. However, due to the simplicity of this construction, we are able to calculate some quantities related to certain spin network states without additional computing power required.

\subsection{Collective indices}
In the generalized construction objects are constructed from tensor products, which increases the amount of indices, so for brevity, and clear analogy to the usual case, we can introduce joint indices, which are the more natural if we vectorize the description. The links of spin \(j\) are equipped with maps:
\begin{equation}
    \hat{\Lambda}(\cdot)\::\:V_j\otimes V_j^*\rightarrow V_j\otimes V_j^*,
\end{equation}
that become matrices \(\hat{\Lambda}\) after admitting the vectorization picture:
\begin{equation}
    \hat{\Lambda}\::\:\mathcal{L}_j\rightarrow\mathcal{L}_j,
\end{equation}
where \(\mathcal{L}_j\cong V_j\otimes V_j^*\), so we can write:
\begin{equation}
    \tensor{\Lambda}{^{a}_{a'}_{b}^{b'}} \equiv \tensor{\Lambda}{^{\mu}_{\nu}}.
\end{equation}
With \(\mu\), \(\nu\) denoting pairs \((a,a')\) and \((b,b')\), respectively. This has to be unpacked when writing explicit contractions or invariants, as the usual indices can be contracted between collective indices \(\mu\), \(\nu\) (as can be seen in \eqref{eq_state_example}), but it can be useful for general index bookkeeping. 

\section{Wilson loop}
\label{Sec:WilsonLoop}

General idea coming from the previous considerations is to extend the unitary holonomy (Wilson line) to the operator \(\hat{\Lambda}\) describing quantum channels (CPTP maps):
\begin{equation}
\hat{h} \rightarrow \hat{\Lambda}. 
\end{equation}
In gauge theories, the most fundamental gauge-invariant object one can construct is the Wilson loop. In the generalized case, the construction remains the same:
\begin{equation}
    W[\hat{\Lambda}]=\mathrm{tr}(\hat{\Lambda}) =  \sum_i\mathrm{tr}\left(\hat{K}_i^*\otimes \hat{K}_i \right). 
\end{equation}

Due to the linearity of the trace and the property that
\begin{equation}
    \mathrm{tr}(\hat{A}\otimes \hat{B})=\mathrm{tr}(\hat{A})\mathrm{tr}(\hat{B}),
\end{equation}
it is also easy to calculate the Wilson loop using results from the usual LQG formulation, again making it more accessible. 
For a unitary case \(\mathrm{tr}(U)=a\in\mathbb{C}\), thus
\begin{equation}
    \mathrm{tr}(\hat{U}^*\otimes\hat{U})=\mathrm{tr}(\hat{U})^*\mathrm{tr}(\hat{U})=|a|^2,
\end{equation}
where we have used that \(\mathrm{tr}(\hat{A}^*)=\mathrm{tr}(\hat{A})^*\), which further implies
\begin{equation}
    W[\hat{\Lambda}]=\mathrm{tr}(\hat{\Lambda}) =  \sum_i \left|\mathrm{tr}(\hat{K}_i)\right|^2. 
\end{equation}
This makes it simple to compare with cases examined in unitary theory. 

Considering the bit flip model, we can calculate the Wilson loop for Kraus operators \eqref{eq_bit_flip_kraus_0} and \eqref{eq_bit_flip_kraus_1}:
\begin{align}
    \begin{split}
        \mathrm{tr}(\hat{\Lambda})&=\sum_i\left|\mathrm{tr}(\hat{K}_i)\right|^2\\
        &=\left|\sqrt{1-p}\cdot2\right|^2+0=4(1-p),
    \end{split}
\end{align}
which smoothly converges to a unitary case of this model as \(p\rightarrow0\), (zero probability of bit flip - identity holonomy) where \(\mathrm{tr}(\hat{\Lambda}=\hat{\mathbb{I}}\otimes\hat{\mathbb{I}})=4\).

\section{Generalized spin network states}
\label{Sec:GeneralizedNetworks}

Spin networks are generalization of the concept of 
Wilson loops, which guarantee gauge-invarince. Spin network states are defined as as square integrable 
functions of the form:
\begin{equation}
    \Psi[A] :=\psi(h_{e_1}[A],\dots,h_{e_E}[A]),
\end{equation}
which are elements of \(L_2\left[SU(2)^E,\prod_ed\mu_H^{(e)}\right]\) space with \(d\mu_H^{(e)}\) being the Haar measure for link \(e\), and \(E\) the number of edges in a given spin network. In this work we aim to formulate a consistent theory of functions of CPTP maps \(\hat{\Lambda}\):
\begin{equation}
    \Psi[A] :=\psi(\hat{\Lambda}_{e_1}[A],\dots,\hat{\Lambda}_{e_E}[A]).
    \label{GeneralizedPsi}
\end{equation}
The projection of functions to gauge-invariant subspace is performed in 
the same manner as in construction of the usual LQG \cite{doná2010introductorylecturesloopquantum}. Namely, at each vertex, group averaging is performed via projection operators with vectorized conjugate action of \(SU(2)\):
\begin{equation}
    \mathcal{P}_{\rm inv}=\int_{SU(2)}dg\bigotimes_eD^*(g)\otimes D(g),
\end{equation}
where the product is performed over all edges connected to a given vertex. 
This procedure is equivalent to imposing the Gauss constraint. 

The projection procedure is discussed in more details in subsection 
\ref {DipoleSpinNetwork}, on example of the dipole spin network. 

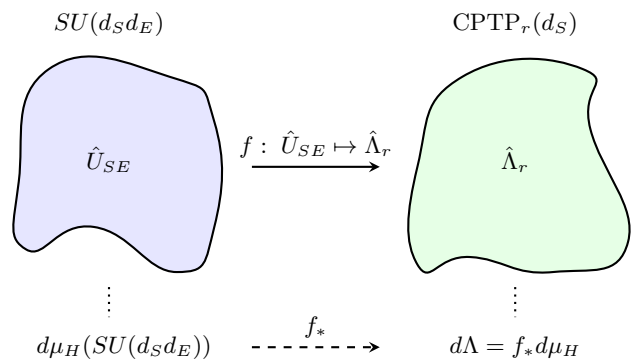
\begin{figure}[H]
    \centering
    \vspace{0.25cm}
    \begin{tikzpicture}[>=stealth, thick]
        \node at (-1.8,2.1) {$SU(d_Sd_E)$};
        \draw[fill=blue!10, smooth cycle, tension=0.9]
          plot coordinates {
            (-3.0,0.0)
            (-2.6,1.5)
            (-1.0,1.4)
            (-0.4,0.8)
            (-0.4,-0.6)
            (-0.9,-1.2)
            (-2.0,-0.6)
            (-2.9,-0.9)
          };
        \node at (-1.8,0.3) {$\hat{U}_{SE}$};
        \node at (3.6,2.1) {$\mathrm{CPTP}_r(d_S)$};
        \draw[fill=green!10, smooth cycle, tension=0.9]
          plot coordinates {
            (2.3,0.2)
            (2.9,1.4)
            (4.5,1.5)
            (4.6,0.7)
            (5.1,-0.8)
            (4.1,-1.2)
            (3.0,-1.0)
            (2.3,-1.1)
          };
        \node at (3.6, 0.3) {$\hat{\Lambda}_r$};
        \draw[->] (0.1,0.2) -- node[above] {$f:\;\hat{U}_{SE}\mapsto\hat{\Lambda}_r$} (1.8,0.2);
        \node at (-1.6,-2.2) {$d\mu_{H}(SU(d_Sd_E))$};
        \node at (3.6,-2.2) {$d\Lambda=f_* d\mu_H$};
        \draw[->, dashed] (0.1,-2.2) -- node[above] {$f_*$} (1.8,-2.2);
        \draw[dotted] (-1.8,-1.4) -- (-1.8,-1.8);
        \draw[dotted] (3.6,-1.4) -- (3.6,-1.8);
        \end{tikzpicture}
    \caption{A diagram of mappings used in our formulation. }
    \label{fig:MappringMeasures}
\end{figure}

The generalized functions \eqref{GeneralizedPsi} are not elements of the 
\(L_2\) space over copies of \(SU(2)\) with Haar measure, but they are able to inherit the measure from 
the joint space of our system and the environment (see Fig. \ref{fig:MappringMeasures}). The total dimension of said joint space is \(\mathrm{dim}(S)\cdot\mathrm{dim}(E) \coloneq d_Sd_E\), where dimension of the system \(d_S\) corresponds to the representation we are considering, and dimension of the environment \(d_E\) depends on the open system model we choose or a coarse graining structure. We will denote elements of this space \(\hat{U}_{SE}\in SU(d_Sd_E)\) and assume that they belong to to special unitary group so that our description can regain unitary case for \(d_E=1\) (trivial environment). The function mapping the total space to our system of interest, and thus description via \(\hat{\Lambda}\) operators is:
\begin{equation}
    f(\hat{U}_{SE})=\sum_i\hat{K}_i^*\otimes \hat{K}_i\equiv\hat{\Lambda}_r,
    \label{eq_manifolds_map}
\end{equation}
which we write in this way to underline that such mapping can be treated as a mapping between two manifolds. 

We assign index \(r\) to the \(\hat{\Lambda}\) operator to distinguish between its ranks understood as the amount of linearily independent Kraus operators it is constructed from (or simply rank of the corresponding Choi matrix), with \(r=1\) corresponding to unitary evolution. This is because the total space of CPTP maps is not a smooth manifold, whereas such maps with fixed rank are, and we will not consider transformations between distinct ranks.

We denote space of CPTP maps with rank \(r\) as \(\mathrm{CPTP}_r(d_S)\). Since \(SU(d_Sd_E)\) is equipped with a Haar measure and we have a well-defined function:
\begin{equation}
f\::\:SU(d_Sd_E)\rightarrow\mathrm{CPTP}_r(d_S),
\end{equation}
we can define a pushforward measure  \(f_*\mu_H\) that lets us perform integration on the latter manifold. It is defined as:
\begin{equation}
    d\Lambda(\lambda)\equiv f_*\mu_H(\lambda)\coloneq d\mu_H(f^{-1}(\lambda)),\:\lambda\subset\mathrm{CPTP}_r,
\end{equation}
where \(f^{-1}\) is to be understood as the preimage of \(f\). Then, for a purpose of defining a scalar product, which will require integrating over the entirety of \(\mathrm{CPTP}_r(d_S)\) space, we can utilize the Haar measure on entirety of \(SU(d_Sd_E)\), as for function \eqref{eq_manifolds_map}, the whole space is the preimage. Explicitly, the space of \(\hat{\Lambda}_r\) operators can be given by:
    \begin{align}
    \begin{split}
    {\rm CPTP}_{r} = &\{\hat{\Lambda}\ :\ \mathcal{B}(\mathcal{H}^S_s)_{\geq0}\rightarrow\mathcal{B}(\mathcal{H}^S_t)_{\geq0}\ \Big|\\&(\hat{\Lambda}\otimes\hat{\mathbb{I}}_n)\geq0\: \forall n \in \mathbb{N} , \\
    &{\rm tr}(\hat{\Lambda}(\hat{\rho}_s))={\rm tr}(\hat{\rho}_s)\:\forall \hat{\rho}_s \in \mathcal{B}(\mathcal{H}^S_s)_{\geq0} , \\
    &\mathrm{dim}(\mathrm{span}(\{\hat{K}_i\})=r\},
    \end{split}
    \end{align}
where \(\mathcal{B}(\mathcal{H}^S_{s/t})_{\geq0}\) are the source/target spaces of density matrices within the system. Relating the measure of the space to the Haar measure on total space lets us compute scalar products of functions over generalized spin networks via the fact that \(\hat{\Lambda}\) oprators are constructed from \(\hat{K}_i\) - Kraus operators related to the space of \(\hat{U}_{SE}\) by:
\begin{equation}
    \hat{K}_i= \bra{i_E}\hat{U}_{SE}\ket{0_E},
\end{equation} 
as introduced in \eqref{eq_Kraus_definition}. In consequence, the integration can eventually be related to integration 
over the space of unitary operators $\hat{U}_{SE}$, with the Haar measure. 
So that: 
\begin{equation}
d\Lambda \coloneq \prod_e d\Lambda_{r_e}  = \prod_e  d\mu_H^{(e)}(\hat{U}^{(e)}_{SE}),
\end{equation}
where the Haar measure is normalised as follows: 
    \begin{equation}
        \int_{SU(d_Sd_E)}d\mu^{(e)}_H(\hat{U}^{(e)}_{SE})=1,
    \end{equation}  
and the pushforward of a measure is guaranteed to be normalized as well. Let us emphasize that choice of this measure is not unique but provides
natural generalization of the Ashtekar-Lewandowski measure,
with a proper unitary limit. Furthermore, by associating 
the space of the $\hat{U}_{SE}$ operators, with the 
special unitary group, the normalizability property can be 
satisfied in a straightforward manner. Utilizing this construction, the kinematical Hilbert space is:
\begin{equation}
\mathcal{H}_{\rm kin} = L^2 \left( \prod_e {\rm CPTP}_{r_e},\prod_e d\Lambda_{r_e}\equiv\prod_e  d\mu_H^{(e)}(\hat{U}^{(e)}_{SE}) \right),
\end{equation}
with scalar product:
\begin{equation}
\braket{\Psi_{\Gamma}|\Phi_{\Gamma'}} :=  \int \prod_{e\in \Gamma \cap \Gamma'}  d\Lambda_{r_e} \overline{\Psi_{\Gamma}[A]} \Phi_{\Gamma'}[A],
\label{ScalarProductNew}
\end{equation}
where \(\Gamma\), \(\Gamma'\) are graphs on which the functions are defined.

\subsection{Single link norm}
     
    For simplicity, consider the expansion of a function of one link with fixed representation:
    \begin{equation}
        \Psi(\Lambda)=\sum_{\mu,\nu}\tensor{\psi}{_{\mu}^{\nu}}\tensor{\Lambda}{^{\mu}_{\nu}}.
        \label{eq_cptp_function_expansion}
    \end{equation}
    For calculating the norm of such a function or a scalar product of two functions, we use Eq. \eqref{ScalarProductNew}. Using collective indices for brevity, we can write:
    \begin{widetext}
    \begin{align}
        \begin{split}
            &\lVert\Psi\rVert^2= \int d\Lambda_r\left\lvert\sum_{\mu,\nu}\tensor{\psi}{_{\mu}^{\nu}}\tensor{(\Lambda_r)}{^{\mu}_{\nu}}\right\rvert^2=\int d\Lambda_r\left\lvert\sum_{\mu,\nu}\tensor{\psi}{_{\mu}^{\nu}} \tensor{\left(\sum_i\hat{K}_i^*\otimes \hat{K}_i\right)}{^{\mu}_{\nu}} \right\rvert^2\\
             &=\int d\mu_H(\hat{U}_{SE})\left\lvert\sum_{\mu,\nu}\tensor{\psi}{_{\mu}^{\nu}}\tensor{\left(\sum_i\mybra{i}{E}\hat{U}_{SE}\ket{0}^*_E\otimes\mybra{i}{E}\hat{U}_{SE}\ket{0}_E\right)}{^{\mu}_{\nu}}\right\rvert^2\\
             &=\int d\mu_H(\hat{U}_{SE})\left\lvert\sum_{\mu,\nu}\tensor{\psi}{_{\mu}^{\nu}}\sum_i\left[\mybra{i}{E}\hat{U}_{SE}\ket{0}^*_E\right]^\mu\left[\mybra{i}{E}\hat{U}_{SE}\ket{0}_E\right]_\nu\right\rvert^2.
        \end{split}
    \end{align}
    \end{widetext}
    This allows us to compute the norm of a function with respect to an arbitrary number of Kraus operators. This approach is not in practice calculable analytically, but due to the Strong Law of Large Numbers, for an integrable function \(f\), and a uniform distribution over the Haar measure \(dU\) over group \(G\), the limit:
    \begin{equation}
         \frac{1}{N}\sum_n^Nf(U^{(n)}) \xrightarrow[N \to \infty]{} \int_GdUf(U),
    \end{equation}
    can be evaluated utilizing the Monte Carlo (MC) integration. Here, \(f(U^{(n)})\) is the \(n\)-th probed element \(U\in G\).
    
    An example of the MC integration will be shown in the following subsection. In the Monte Carlo integration matrices \(U_{SE}(d_Sd_E)\) are valid only for minimal Kraus construction, i.e., all Kraus operators being linearly independent so that \(d_E=r\). Otherwise, \(d_E\) should be replaced with the rank \(r\) of \(\hat{\Lambda}_r\) operator when relating the integration to overcomplete Kraus set.
    For completeness, let us explicitly write the product of two functions in the extended theory, utilizing the Einstein summation convention:
    \begin{widetext}
    \begin{align}
        \langle\Psi |\Phi\rangle&=\int d\Lambda_r\psi^*(\Lambda_r)\phi(\Lambda_r)=\int d\Lambda_r\tensor{\psi}{^*_\mu^{\mu'}}\tensor{\phi}{_\nu^{\nu'}}\tensor{(\Lambda_r)}{^{*\mu}_{\mu'}}\tensor{(\Lambda_r)}{^\nu_{\nu'}}
        =\int d\Lambda_r\tensor{\psi}{^*_\mu^{\mu'}}\tensor{\phi}{_\nu^{\nu'}}\sum_{i,j}\tensor{(K_i\otimes K_i^*)}{^\mu_{\mu'}}\tensor{(K_j^*\otimes K_j)}{^\nu_{\nu'}}\nonumber\\
        &=\int d\mu_H(\hat{U}_{SE})\tensor{\psi}{^*_\mu^{\mu'}}\tensor{\phi}{_\nu^{\nu'}}\sum_{i,j}\left[\mybra{i}{E}\hat{U}_{SE}\ket{0}_E\right]^\mu\left[\mybra{i}{E}\hat{U}_{SE}\ket{0}^*_E\right]_{\mu'}\left[\mybra{j}{E}\hat{U}_{SE}\ket{0}^*_E\right]^\nu\left[\mybra{j}{E}\hat{U}_{SE}\ket{0}_E\right]_{\nu'},
    \end{align}
    \end{widetext}
    which again can be calculated via the Monte Carlo sampling of \(\hat{U}_{SE}\).
    \subsection{Wilson loop norm}
    As an example of practical calculation of the constructed scalar product, we present results for the norm of the Wilson loop in the fundamental representation \(j=1/2\):
    \begin{equation}
        \Psi=\mathrm{tr}(\hat{\Lambda}_{r}),
        \label{eq_wilson_loop_function}
    \end{equation}
    and comparing them with analytical results for the case of unitary \(\hat{\Lambda}\). Square of norm of the Wilson loop yields:
    \begin{equation}
        \lVert\Psi\rVert^2=\braket{\Psi|\Psi}=\int d\Lambda_r|\mathrm{tr}(\hat{\Lambda}_r)|^2.
        \label{eq_wilson_loop_norm_sq}
    \end{equation}
    Considering a case with only one Kraus operator, therefore a unitary evolution
    (corresponding to rank $r=1$), the expression for the norm of a Wilson loop simplifies and can be computed analytically through the characters of representations. In the unitary case:
    \begin{equation}
        \hat{\Lambda}_1 = \hat{U}^*\otimes \hat{U},
    \end{equation}
    and as discussed earlier, calculating the trace simplifies to
    \begin{equation}
        \mathrm{tr}(\hat{U}^*\otimes \hat{U})=\mathrm{tr}(\hat{U})^*\mathrm{tr}(\hat{U}).
    \end{equation}
    Since \(U\) is an irrep in this case, its trace is the representation character. Since we are considering a fundamental representation, we have:
    \begin{equation}
        \mathrm{tr}(\hat{U})=\chi_{1/2}(\hat{U}),
    \end{equation}
    and we can use the property that:
    \begin{equation}
        \chi_{1/2}(\hat{U})\chi_{1/2}(\hat{U})=\chi_0(\hat{U})+\chi_1(\hat{U}),
    \end{equation}
    as well as the orthogonality of characters:
    \begin{equation}
   \braket{\chi_i|\chi_j}  =    \int_Gdg\chi_i^*(\hat{U}(g))\chi_j(\hat{U}(g))=\delta_{ij},
    \end{equation}
    to calculate the square of the norm:
    \begin{align}
        \begin{split}
            &\lVert\mathrm{tr}(\hat{\Lambda}_1)\rVert^2=\int d\mu_H(\hat{U})\left|\mathrm{tr}(\hat{\Lambda}_1)\right|^2=\int d\mu_H(\hat{U})\left|\mathrm{tr}(\hat{U})\right|^4\\
            &=\int d\mu_H(\hat{U})\left|\chi_{1/2}\right|^4=\int d\mu_H(\hat{U})\chi_{1/2}^2\chi_{1/2}^{*2}\\
            &=\underbrace{\braket{\chi_{0}|\chi_0}}_{1}+\underbrace{\braket{\chi_0|\chi_1}}_{0}+\underbrace{\braket{\chi_1|\chi_0}}_{0}+\underbrace{\braket{\chi_1|\chi_1}}_{1}
            =2,
        \end{split}
        \label{eq_unitary_wilson_loop_norm}
    \end{align}
    where we have omitted arguments of the characters \(\chi_i :=\chi_i(\hat{U})\). \par
\begin{figure}[H]
        \centering
        \includegraphics[width=\linewidth]{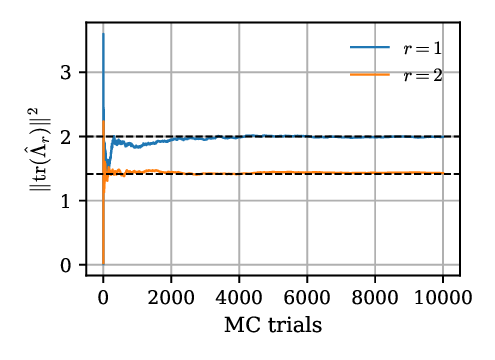}
        \caption{Results of Monte Carlo integration of Wilson loop norm of \(\hat{\Lambda}_r\) for \(r=1\) corresponding to unitary evolution and for \(r=2\) for rank 2 environment, which encapsulates, e.g., the bit flip model considered earlier. Both results are considered for a fundamental representation, i.e., 2-dimensional system corresponding to the fundamental representation. Obtained values converge within a feasible number of Monte Carlo trials. For the unitary case results coincide with the theoretical calculations. Horizontal dashed lines are drawn for values 2 and \(\sqrt{2}\).}
        \label{fig_MC_wilson_loop_norm}
    \end{figure}

 For the cases of $r=1$ and $r=2$, performing Monte Carlo calculations of squared norm of the Wilson loop function \eqref{eq_wilson_loop_norm_sq} yields the results shown in Fig. \ref{fig_MC_wilson_loop_norm}.

    The MC integaration confirms the analitically obtained value of  $\lVert\mathrm{tr}(\hat{\Lambda}_1)\rVert^2=2$. In turn, for the 
    non-unitary quantum channel with $r=2$, the $\lVert\mathrm{tr}(\hat{\Lambda}_2)\rVert^2\approx \sqrt{2}$ is predicted. 

\begin{figure}[H]
        \centering
        \includegraphics[width=\linewidth]{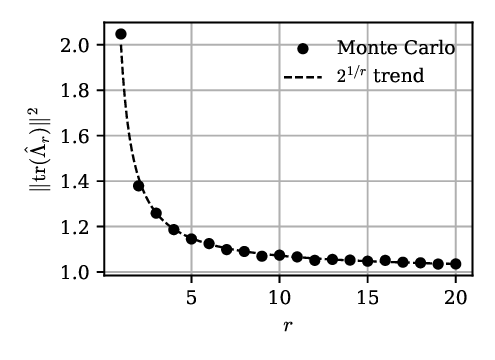}
        \caption{Relation between square of the norm of Wilson loop and rank of the \(\hat{\Lambda}_r\) operator. Computed values follow a \(2^{1/r}\) trend. As the degrees of freedom of environment increase, the probability gets more diluted. Results obtained for 10000 samples for each point.}
        \label{fig:MCnormR120}
    \end{figure}

   The numerically obtained values of $\lVert\mathrm{tr}(\hat{\Lambda}_r)\rVert^2$ for $r \in \{1,2,...,20\}$ are shown in Fig. \ref{fig:MCnormR120}.
   The MC results suggest that the norm satisfies relation $\lVert\mathrm{tr}(\hat{\Lambda}_r)\rVert^2=2^{1/r}$, which remains to be proven analytically.

\subsection{Dipole spin network}
\label{DipoleSpinNetwork}

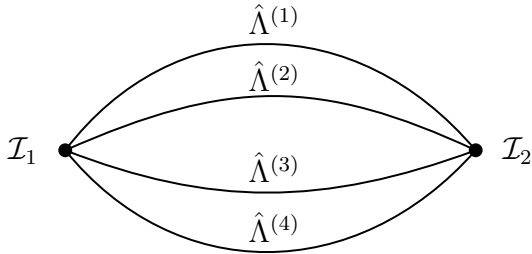
\begin{figure}[H]
    \centering
    \large
    \tikzset{every picture/.style={line width=0.75pt}}
        \begin{tikzpicture}[x=0.75pt,y=0.75pt,yscale=-1]
        \coordinate (L) at (38.5,86.5);
        \coordinate (R) at (245.5,86.5);
        \fill (L) circle (3.5);
        \fill (R) circle (3.5);
        \draw (L) .. controls (93,15)  and (185,15)  .. (R);
        \draw (L) .. controls (121,50) and (167,50) .. (R);
        \draw (L) .. controls (112,115) and (168,115) .. (R);
        \draw (L) .. controls (94,155) and (185,155) .. (R);
        \node at (144,20)  {$\hat{\Lambda}^{(1)}$};
        \node at (144,49)  {$\hat{\Lambda}^{(2)}$};
        \node at (144,95)  {$\hat{\Lambda}^{(3)}$};
        \node at (144,125) {$\hat{\Lambda}^{(4)}$};
        \node[left=6pt]  at (L) {$\mathcal{I}_{1}$};
        \node[right=6pt] at (R) {$\mathcal{I}_{2}$};
        \end{tikzpicture}
    \caption{A dipole spin network state in the generalized description.}
    \label{Fig.DipoleNetwork}
\end{figure}

The simplest example relevant to a 3D space is a spin network consisting of 
two nodes and four links between them (as shown in Fig. \ref{Fig.DipoleNetwork}),
corresponding to two tetrahedra sharing all of the faces. Below, an example of a gauge invariant construction for this case is provided.

Here, \(\mathcal{I}_i\in\mathrm{Inv}\left[{(V_{1/2}\otimes V_{1/2}^*)}^{\otimes4}\right]\), with mappings \(\hat{\Lambda}^{(i)} \in V_{1/2}\otimes {V^*_{1/2}}\otimes{V^*_{1/2}}\otimes V_{1/2}\), thus in general a dipole state can be expressed e.g. as:
\begin{align}
    \begin{split}
        \Psi =\;& \tensor{\mathcal{I}}{_1^{m_2}_{m_1}^{m_4}_{m_3}^{m_6}_{m_5}^{m_8}_{m_7}} \\
        & \tensor{\Lambda}{_{1}^{m_1}_{m'_1}_{m_2}^{m'_2}}
          \tensor{\Lambda}{_{2}^{m_3}_{m'_3}_{m_4}^{m'_4}}
          \tensor{\Lambda}{_{3}^{m_5}_{m'_5}_{m_6}^{m'_6}}
          \tensor{\Lambda}{_{4}^{m_7}_{m'_7}_{m_8}^{m'_8}} \\
        & \tensor{\mathcal{I}}{_2^{m'_1}_{m'_2}^{m'_3}_{m'_4}^{m'_5}_{m'_6}^{m'_7}_{m'_8}},
    \end{split}
    \label{eq_state_example}
\end{align}
but other contractions are also possible. The invariant spaces for 8 spins \(j=1/2\) are 14 dimensional at each vertex, so there are \(14^2=196\) contributing states. Below, we give a representative state from this pool:
\begin{align}
    \begin{split}
        \Psi =\;& 
        \tensor{\delta}{^{m_2}_{m_1}}
        \tensor{\delta}{^{m_4}_{m_3}}
        \tensor{\delta}{^{m_6}_{m_5}}
        \tensor{\delta}{^{m_8}_{m_7}} \\
        & 
        \tensor{\Lambda}{_{1}^{m_1}_{m'_1}_{m_2}^{m'_2}}
        \tensor{\Lambda}{_{2}^{m_3}_{m'_3}_{m_4}^{m'_4}}
        \tensor{\Lambda}{_{3}^{m_5}_{m'_5}_{m_6}^{m'_6}}
        \tensor{\Lambda}{_{4}^{m_7}_{m'_7}_{m_8}^{m'_8}} \\
        & 
        \tensor{\delta}{^{m'_1}_{m'_2}}
        \tensor{\delta}{^{m'_3}_{m'_4}}
        \tensor{\delta}{^{m'_5}_{m'_6}}
        \tensor{\delta}{^{m'_7}_{m'_8}}.
    \end{split}
\end{align}

Note that here, we used \(\delta\) tensors as invariant building blocks instead of usual \(\epsilon\) tensors. This will be discussed in the next section.

\section{General construction}
\label{Sec:GeneralConstruction}

In the extended description of LQG consisting of density matrices and quantum channels that was introduced throughout this paper, an arbitrary state can be constructed using the same approach as in the unitary case. LQG states are obtained by contracting representation matrices of mappings carried by the links via intertwiners at the nodes. Intertwiners are in general elements of \(SU(2)\)-invariant subspace of basis elements of \(L_2\) space over the spin network. In our description, we have to adapt their form to the generalized mappings we introduce. At a given vertex, we have:
\begin{equation}
    \mathcal{I}_v\in\mathrm{Inv}_{SU(2)}\Big(\bigotimes_e\mathcal{L}_{j_e}\Big).
\end{equation}
The group acts on a Liouville space \(\mathcal{L}\cong V\otimes V^*\) by the adjoint action, so the invariant tensors change accordingly. To pair two spaces into a singlet now we will use the \(\delta\indices{^a_b}\in\mathcal{L}_{1/2}\) tensor as opposed to \(\epsilon\) in the case of two (non-)dual spaces:
\begin{equation}
    {\delta'}\indices{^a_b}=g\indices{^a_c}\delta\indices{^c_d}(g^{-1})\indices{^d_b}=g\indices{^a_c}(g^{-1})\indices{^c_b}=\delta\indices{^a_b}.
\end{equation}

For the construction of higher-dimensional invariants, a combination of delta tensors can be used. The use of epsilon tensors is still allowed as they can be decomposed into delta tensors via the indentity:
\begin{equation}
    \epsilon^{ab}\epsilon_{cd}=\delta\indices{^{c}_{a}}\delta\indices{^{d}_{b}}-\delta\indices{^{d}_{a}}\delta\indices{^{c}_{b}},
\end{equation}
but they do not introduce any linearly independent invariants.

The intertwiners are thus operators invariant under the gauge transformation, or, equivalently, by elements that commute with 
the group action:
\begin{equation}
\mathcal{I}_v=\hat{\mathcal{U}}_v\mathcal{I}_v\hat{\mathcal{U}}_v^\dagger.
\end{equation}

Writing down a spin network state is the same as in the usual formulation: take all holonomy operators and contract them with intertwiners:
\begin{equation}
    \Psi=\prod_e\hat{\Lambda}(j_e)\prod_v\mathcal{I}_v.
\end{equation}

This is a minimal construction for introducing open systems into the LQG formalism, which corresponds to tracing out part of the spin network while allowing it to influence the considered part (the system) via quantum channels.

As presented in \cite{Curtright_2017} and proven in detail in the appendix of \cite{Mendon_a_2013}, the general formula for finding the multiplicity of spin \(s\) subspace for a product of \(n\) spins \(j\) is given by: 
\begin{equation}
    \left[(V_j^{\otimes n})\right]_s=c_0(s,n,j)-c_0(s+1,n,j),
\end{equation}
where:
\begin{widetext}
    \begin{equation}
    c_0(s,n,j)=\sum_{k=0}^{\left|\frac{nj+s}{2j+1}\right|}(-1)^k{n\choose k}{nj+s-(2j+1)k+n-1\choose nj+s-(2j+1)k}.
\end{equation}
\end{widetext}

The above expression holds for any number and order or dual/non-dual spaces included, as follows from the isomorphism between a space and its dual. For the case of \(n\) spins \(j=1/2\) and the (invariant) subspace of spin \(s=0\), this formula reduces to:
\begin{equation}
    \mathrm{dim}\left(\mathrm{Inv}(V_{1/2}^{\otimes n})\right)=
    \begin{cases}
        0\quad&\text{for}\quad n\text{ - odd}\\
        \frac{1}{1+n/2}{n\choose n/2}\quad&\text{for}\quad n\text{ - even}
    \end{cases},
\end{equation}
where the formula for an even number of spins is the formula for the \(m=n/2\)-th \emph{Catalan number}. The latter can be represented in various ways, e.g. \emph{Dyck words}, which might help in generating all linearily independent contractions for a large multiplicity of spaces.

\subsection{Density matrix of a full spin network}
Findings presented in the previous sections of this paper provide a simple extension and, while computationally more expensive, could still be within the reach of current technology and possibly useful for exploring the nature of spacetime. However, a more general construction of open loop quantum gravity is possible at the level of formulation of the theory while introducing the cylindrical functions as the gauge and diffeomorphism invariant states of the kinematical Hilbert states (see e.g. section 3 of \cite{doná2010introductorylecturesloopquantum}). The density matrices can be introduced as mappings on the space of these functions, allowing of mixing of geometries of spin networks and the spins within them. Peter-Weyl theorem tells us that (in particular) \(L_2\) spaces can be decomposed into the basis constructed from a direct sum of tensor products of irreducible vector spaces, so a mapping on such a space, for a single \(SU(2)\) copy would be
\begin{equation}
   \hat{\rho} \in  \mathcal{B}(\mathcal{H}_{\rm kin}) =\bigoplus_j\left(V_j\otimes V_j^*\right)\otimes\bigoplus_k\left(V_k\otimes V_k^*\right)^*.
\end{equation}
Using \((V_k*)^*=V_k\) and the fact that the tensor product is distributive over direct sum up to an isomorphism, we get:
\begin{align}
    \hat{\rho}&\in\bigoplus_{j,k}\left(V_j\otimes V_j^*\right)\otimes\left(V_k^*\otimes V_k\right)\nonumber \\
    &\cong\bigoplus_{j,k}\left(V_j\otimes V_j^*\otimes V_k^*\otimes V_k\right)\nonumber \\
    &\cong\bigoplus_{j,k}\left(\underbrace{V_j\otimes V_k^*}_{V_T}\otimes \underbrace{V_j^*\otimes V_k}_{V_T^*}\right),
\end{align}
where we have used associativity and commutativity up to isomorphic properties. For the whole spin network \(SU(2)^L\) with \(L\) edges we generalize this to:
\begin{align}
    \hat{\rho}&\in\left[\bigoplus_{j_1,...j_L}\bigotimes_l^L\left(V_{j_l}\otimes V_{j_l}^*\right)\right]\otimes\left[\bigoplus_{j'_1,...j'_K}\bigotimes_k^K\left(V_{j'_k}^*\otimes V_{j'_k}\right)\right]  \nonumber \\
    &\cong\bigoplus_{\substack{j_1...j_L  \\ j'_1...j'_K}}\bigotimes_{l,k}^{L,K}\left(V_{j_l}\otimes V_{j_l}^*\right)\otimes \left(V_{j'_k}^*\otimes V_{j'_k}\right),
\end{align}
which can again be reshuffled, like in the case of one link. The dimension of spaces in this approach grows rapidly, and it is probably infeasible to consider such systems in the current state of technology and knowledge about LQG. Therefore, the approach proposed previously in this paper gives a more promising tool for describing the noisy spacetime structure.

\section{Summary and outlook}
\label{Sec.Summary}

In this article, we have proposed a generalization of 
gauge-invariant spin network states in which the $SU(2)$ 
holonomies are replaced by quantum channels. We have shown 
that gauge-invariance can be preserved even when departing 
from unitary maps. This allowed us to construct generalized 
spin-network states expressed in terms of the Kraus operators. 
Furthermore, we have proposed an inner product, which allows 
us to construct assiciated Hilbert space. In the unitary limit, 
the standard construction of the Hilbert space of spin networks
is recovered. 

These results open new perspectives for the study of open quantum 
gravitational systems and other $SU(2)$ gauge-invariant systems,
interacting with environment. 

The discussion in this article was restricted to the construction 
of states and the corresponding Hilbert space. Several important 
directions for future work naturally arise, including:
\begin{itemize}
\item the action of the generalized non-unitary holonomies 
and fluxes on the states introduced here;
\item a generalization of the holonomy-flux algebra;
\item the definition and analysis of geometric operators, 
such as area and volume;
\item the investigation of the resulting quantum dynamics.
\end{itemize}

Finally, let us emphasize that extending holonomies to quantum channels 
may introduce additional degrees of freedom. Consequently, the 
reconstructed semiclassical theory may contain degrees 
of freedom beyond those of pure gravity. It is, therefore, natural 
to explore whether such additional structures could be interpreted 
as an emergent matter sector. Potential applications to cosmology 
and black hole thermodynamics are also of interest, as they may provide 
possible testing grounds for the ideas developed here. 

\section*{Appendix: Kraus operators}
\label{Sec.Appendix}

Here, we show how Kraus operators emerge from constraining
ourselves to describe only a part of the system, and comment 
on some of their properties.

Kraus operators canonically describe the evolution of a state within the 
system is subject to the influence of some environment. The 
dynamics of the system are then described by the partial 
trace over the environment degrees of freedom:
\begin{align}
    \begin{split}
        \hat{\rho}_S(t)&={\rm tr}_E(\hat{\rho}_{SE}(t))\\
        &=\sum_{i_E}^{d_E}\bra{i_E}\hat{U}_{SE}(\hat{\rho}(0)\otimes\hat{\rho}_E)\hat{U}_{SE}^\dagger\ket{i_E},
    \end{split}
\end{align}
where \(d_E\) is the dimension of the environment's Hilbert 
space, and \(i_E\) denote its basis vectors. Owing to the 
fact that we can freely expand the environment so that  \(\hat{\rho}_E\) is purified, we can choose the basis such that:
\begin{equation}
    \hat{\rho}_E=\ket{0_E}\bra{0_E}.
\end{equation}

This lets us write:
\begin{align}
    \begin{split}
        \hat{\rho}_S(t)&={\rm tr}_E(\hat{\rho}_{SE}(t))\\
&=\sum_{i_E}\bra{i_E}\hat{U}_{SE}\big(\hat{\rho}_S(0)\otimes\ket{0_E}\bra{0_E}\big)\hat{U}_{SE}^\dagger\ket{i_E},
    \end{split}
\end{align}
and  by acting with \(\hat{U}_{SE}\) on state \(\ket{0_E}\), we find:
\begin{align}
\hat{\rho}_S(t)&=\sum_{i_E}\bra{i_E}\hat{U}_{SE}\ket{0_E}
\hat{\rho}_S(0)\bra{0_E}\hat{U}_{SE}^\dagger\ket{i_E}\\
    &=\sum_{i}\hat{K}_i\hat{\rho}_S(0)\hat{K}_i^\dagger,
    \label{eq_kraus_evolution}
\end{align}
where we have introduced \textit{Kraus operators}:
\begin{equation}
    \hat{K}_i(t):= \bra{i_E}\hat{U}_{SE}(t)\ket{0_E}.
\end{equation} 

Note that their explicit form will depend on the choice of 
basis, but the whole evolution is equivalent up to a unitary
transformation, i.e., change of basis:
\begin{equation}
    \hat{K}_i'=\sum_l\tensor{U}{_i^l}\hat{K}_l,
\end{equation}
so that:
\begin{align}
    \begin{split}
        \hat{\rho}_t&=\sum_i\hat{K}'_i\hat{\rho}_s\hat{K}_i^{'\dagger}=\sum_i\sum_{l,l'}\tensor{U}{_i^l}\hat{K}_l\hat{\rho}_s\tensor{U}{^*_i^{l'}}\hat{K}_{l'}^{\dagger}\\
        &=\sum_{l,l'}\underbrace{\sum_i\tensor{U}{_i^l}\tensor{U}{^*_i^{l'}}}_{\delta^{ll'}}\hat{K}_l\hat{\rho}_s\hat{K}_{l'}^{\dagger}=\sum_l\hat{K}_l\hat{\rho}_s\hat{K}_l^\dagger,
    \end{split}
\end{align}
where we have used the unitarity of \(\hat{U}\) to obtain the Kronecker delta \(\delta_{ll'}\). The Kraus operators are in general \textit{some} complex operators, but they do not have to be (and if they are describing non-unitary mappings - they are not) unitary, but they are all trace non-increasing:
\begin{equation}
    \sum_i\hat{K}^\dagger_i\hat{K}_i\leq\hat{\mathbb{I}},
\end{equation}
where the inequality is saturated for a pure-state environment, which is the case we consider here. The condition is to be understood as:
\begin{equation}
    \hat{\mathbb{I}}-\sum_i\hat{K}_i^\dagger \hat{K}_i\geq0,
\end{equation}
i.e., the difference between the identity map and the sum of the products of the Kraus operators is positive semidefinite, thus being unable to increase the total probability of states it acts on. The operators can, however, be constructed in such a way that they decrease the total probability, like in a case of post-selection of measurement results or leak of probability to observables that we cannot observe. These cases will not be discussed here.
Notice the hermitian conjugate operator is on the left here as opposed to the evolution equation \eqref{eq_kraus_evolution}.

Let us now consider changing the basis in the source space 
\(\mathcal{H}_s\) and the target space \(\mathcal{H}_t\), 
i.e., gauge transformation:
\begin{align}
    \begin{split}
        \hat{\rho}'_t&=\sum_i\hat{K}'_i\hat{\rho}'_s \hat{K}_i^{\prime\dagger},\\
        \hat{U}_t\hat{\rho}_t\hat{U}_t^\dagger&
        =\sum_i\hat{K}'_i\hat{U}_s\hat{\rho}_s\hat{U}_s^\dagger \hat{K}_i^{\prime\dagger},
        \label{eq_quantum_channel_holonomy_base_change}
    \end{split}
\end{align}
and since the evolution should not depend on the choice of basis, from this we can infer the transformation properties of the Kraus operators by multipying the expression \eqref{eq_quantum_channel_holonomy_base_change} from the left by \(\hat{U}_t^\dagger\), and from the right by \(\hat{U}_t\) and utilizing the fact that \(\hat{U}_t^\dagger \hat{U}_t=\hat{U}_t\hat{U}_t^\dagger=\hat{\mathbb{I}}\):
\begin{equation}
    \hat{\rho}_t=\sum_i\hat{U}_t^\dagger \hat{K}_i'\hat{U}_s\hat{\rho}_s\hat{U}_s^\dagger \hat{K}_i^{\prime\dagger}\hat{U}_t.
\end{equation}

By noticing that \(\hat{U}_s^\dagger \hat{K}_i^{\prime\dagger}\hat{U}_t=(\hat{U}_t^\dagger \hat{K}_i'\hat{U}_s)^\dagger\) and comparing the formula 
above to \eqref{eq_kraus_evolution} we get that:
\begin{align}
    \begin{split}
        \hat{K}_i\rightarrow \hat{U}_t \hat{K}_i\hat{U}_s^\dagger.
    \end{split}
\end{align}
This is the same transformation rule as for the holonomy \eqref{eq_holonomy_transformation}. 

For CPTP maps \(\hat{\Lambda}(\hat{\rho})=\sum_i\hat{K}_i\hat{\rho} \hat{K}_i^\dagger\) extended to some environment we use the maps \(\hat{\Lambda}\otimes\hat{\mathbb{I}}\) and redefine the Kraus operators to be \((\hat{K}_{SE})_i=\hat{K}_i\otimes\hat{\mathbb{I}}\). 
As can be shown they fulfill the requirement \(\sum_i\hat{K}_i^\dagger \hat{K}_i=\hat{\mathbb{I}}\):
\begin{align}
    \begin{split}   
            &\sum_i(\hat{K}_{SE})_i^\dagger(\hat{K}_{SE})_i=\sum_i(\hat{K}_i\otimes\hat{\mathbb{I}})^\dagger(\hat{K}_i\otimes\hat{\mathbb{I}})\\
            &=\sum_i(\hat{K}_i^\dagger \hat{K}_i\otimes\hat{\mathbb{I}})=\left(\underbrace{\sum_i\hat{K}_i^\dagger \hat{K}_i}_{\hat{\mathbb{I}}}\otimes\hat{\mathbb{I}}\right)=\hat{\mathbb{I}}_{SE}.
    \end{split}
\end{align}

Thus, for the extended map \(\hat{\Lambda}\otimes\hat{\mathbb{I}}\), by definition, the evolution is given by:
\begin{equation}
    (\hat{\Lambda}\otimes\hat{\mathbb{I}})(\hat{\rho})\coloneq\sum_i(\hat{K}_i\otimes\hat{\mathbb{I}})\hat{\rho}(\hat{K}_i^\dagger\otimes\hat{\mathbb{I}}),
\end{equation}
which is the manifestation of the assumption that the environment does not change during the transformation. This could possibly be useful while utilizing Choi-Jamiołkowski isomorphism equivalent for quantum channels.
\bibliography{references}

\end{document}